\documentclass[final,5p,times,twocolumn]{elsarticle}

\usepackage{amsmath,amssymb}
\usepackage{graphicx}
\usepackage{booktabs}
\usepackage{hyperref}

\journal{Physics Letters B}

\AtBeginDocument{%
  \setlength{\bibsep}{0pt plus 0.2ex}%
}

\begin{document}

\begin{frontmatter}

\title{Astrons: Reissner--Nordstr\"om Primordial Naked Singularities}

\author[lecce,cnr]{Claudio Corian\`o}
\author[lecce]{Paul H. Frampton}
\author[lecce]{Leonardo Torcellini}

\address[lecce]{Dipartimento di Matematica e Fisica ``Ennio De Giorgi'',
Universit\`a del Salento and INFN-Lecce, Via Arnesano, 73100 Lecce, Italy}
\address[cnr]{CNR Nanotec, Lecce, Italy}

\begin{abstract}
We summarize a set of constraints on a proposed population of primordial,
ultra-massive, electrically charged compact objects, which we call astrons.
The analysis combines charge generation, charge saturation, persistence of the
charge in an ionized medium, screening by the intergalactic plasma, the
Reissner--Nordstr\"om geometry of highly charged compact objects, lensing, and
the cosmological implications of a sparse charged population.  We also discuss
their possible role as primordial dark seeds for early structure formation,
rather than as luminous objects directly observed at high redshift.  The resulting
scenario is sharply constrained.  Ordinary accretion saturation gives charges
far below the large-charge phenomenological benchmark, screening is a serious
plasma-physics issue, and a large charge can place the exterior geometry deep in
the super-extremal regime.  As expected at the level of a homogeneous
Friedmann--Lema\^{\i}tre--Robertson--Walker (FLRW) description, the interaction
energy of a population of charged objects scales as
\(a^{-4}\), so the simplest perfect-fluid reduction does not generate
asymptotically late-time acceleration; any acceleration era tied to that
homogeneous component can only be transitory.  The astron scenario should be
regarded as a constrained framework whose viability depends on plasma physics
and on a cosmological treatment beyond the homogeneous approximation.
\end{abstract}


\end{frontmatter}

\section{Introduction}

The possibility that dark matter and dark energy are connected to a population
of primordial compact objects has motivated a variety of mechanisms in which
gravity is supplemented by long-range interactions
\cite{Frampton2022EAU,Frampton2023DMDE,BogoradGrahamRamani2025}.  In this
Letter we discuss a specific realization: ultra-massive compact objects carrying
a macroscopic electric charge.  Charged compact objects and charged primordial
black-hole scenarios have been considered in several related contexts
\cite{ZajacekTursunov2019,ZajacekEtAl2018,ArayaEtAl2022}.  We refer to such
objects as \emph{astrons}.  In the phenomenological proposal of
Refs.~\cite{Frampton2022EAU,Frampton2023DMDE}, astrons were introduced as a
sparse population of very massive charged objects whose electrostatic repulsion
could be relevant to cosmic acceleration.  The fiducial large-charge branch
examined here is
\begin{equation}
M_A\sim 10^{12}M_\odot,
\qquad
Q_A\sim 4\times 10^{32}\ {\rm C},
\end{equation}
with inter-object separations of order megaparsecs.

This charge scale is not the output of ordinary late-time accretion.  It should
be read as a primordial or early-universe charge-concentration hypothesis.  The
first question is therefore whether an object of mass \(M_A\) could ever be
formed with a net charge of order \(Q_A\).  If such a charge is assumed to have
been generated, the next question is whether its cosmological effect really
behaves as late-time acceleration, or instead appears only as a transient or
domain-dependent effect once the homogeneous and inhomogeneous descriptions are
distinguished.

The word astron is used here in an operational sense.  It denotes a dark,
ultra-massive charged gravitating source, together with the electromagnetic and
plasma environment needed to support or screen its charge.  It should not be
pictured as an ordinary luminous star.  A realistic object may contain a compact
core, a horizonless charged interior, or a singular electrovac center, and it
may be surrounded by a very dilute, non-luminous gas or plasma envelope on
sub-parsec to parsec scales.  The present Letter does not assume a detailed
interior model; it uses the mass, charge and exterior field as the measurable
inputs.

In this sense an isolated astron would not be ``seen'' because it shines.  It
would be inferred through its gravitational field, through the response of the
surrounding plasma, or through lensing.  The lensing point is subtle.  The mass
still bends light, so weak-field mass lensing remains.  The charge, however,
does not make the optical signal simply stronger.  In the
Reissner--Nordstr\"om exterior the \(Q^2/r^2\) term reduces the next-to-leading
weak-field deflection and, for sufficiently large \(\Xi\), removes the photon
sphere.  Thus a highly charged astron is not expected to look like a more
strongly lensing black hole; it should instead move away from the standard
black-hole strong-lensing class.

The detailed derivations of the capture model, charge saturation, screening,
horizon structure, lensing, homogeneous cosmology and backreaction formulation
will be given elsewhere \cite{CorianoEtAlPrep}.  The discussion below extracts
the physical results and organizes them as consistency conditions on the astron
scenario.

The motivation for considering astrons is simple.  If two objects of mass \(M\)
and charge \(Q\) are separated by a distance \(r\), the ratio of electrostatic
repulsion to Newtonian attraction is
\begin{equation}
\Xi\equiv \frac{k_eQ^2}{GM^2}.
\end{equation}
Equivalently,
\begin{equation}
\frac{F_{\rm em}}{|F_{\rm grav}|}=\Xi ,
\qquad
F_{\rm em}=\frac{k_eQ^2}{r^2},
\qquad
|F_{\rm grav}|=\frac{GM^2}{r^2}.
\end{equation}
The benchmark charge above gives \(\Xi_A\simeq 5.4\).  Thus, at the level of a
pairwise force estimate, the electric repulsion exceeds the gravitational
attraction.  The question is whether such a statement can survive the
astrophysical, geometrical and cosmological consistency tests required of a
model.

This is the point at which the present analysis departs from the earlier
phenomenological motivation in Refs.~\cite{Frampton2022EAU,Frampton2023DMDE}.
Those works emphasized the possibility that a sparse population of massive
charged objects could connect dark matter and cosmic acceleration through
long-range electrostatic repulsion.  Here we keep that benchmark as the starting
charge scale, but we ask where the effect can actually reside.  Pairwise
repulsion is a local dynamical statement.  Cosmic acceleration is a statement
about the averaged expansion.  The same charge must therefore pass the
generation and saturation tests, survive plasma screening, define a consistent
Einstein--Maxwell exterior, produce an acceptable lensing signature, and enter a
cosmological averaging calculation.  The result is a sharpened and more
restrictive version of the proposal, in which the homogeneous contribution is
separated from any possible backreaction effect.

The logic of the Letter is organized around two levels of the problem.  The
first is the local and compact-object level: can a large net charge be produced,
can it persist in an ionized medium, and what exterior geometry and lensing
properties follow from the resulting charge-to-mass ratio?  The second is the
cosmological level: if such sources exist, does their charge produce an
accelerating large-scale expansion after the appropriate averaging procedure?

Keeping these two levels distinct is essential.  At the first level, charge can
respond rapidly during collapse, but ordinary saturation gives a much smaller
charge than the benchmark; the same benchmark charge that gives pairwise
repulsion also removes the photon sphere.  At the second level, the homogeneous
Coulomb energy redshifts like radiation, so cosmic acceleration cannot be
identified with the pairwise force alone.  Any remaining acceleration effect has
to come from a genuine inhomogeneous Einstein--Maxwell backreaction.  The astron
proposal is therefore a chain of conditions, any one of which can become the
limiting step.

\section{Primordial origin and early structure}

The large-charge benchmark should be understood as a primordial hypothesis, not
as the outcome of ordinary late-time accretion.  If an object with
\(M_A\sim10^{12}M_\odot\) and \(Q_A\sim4\times10^{32}\,{\rm C}\) exists, the
charge must have been produced during an early phase of cosmic history, or by a
nonstandard mechanism that precedes normal galaxy formation.

A useful order-of-magnitude guide is the horizon mass in the radiation era,
familiar from primordial black-hole formation \cite{CarrHawking1974},
\begin{equation}
M_H(t)\sim \frac{c^3t}{G}
\simeq
2\times10^5M_\odot
\left(\frac{t}{1\,{\rm s}}\right).
\end{equation}
Thus the fiducial mass \(M_A\sim10^{12}M_\odot\) corresponds to
\begin{equation}
t_A\sim5\times10^6\,{\rm s},
\end{equation}
namely a time of order months after the Big Bang.  This estimate does not
constitute a formation model.  It only indicates that, if astrons are primordial
objects of this mass, their origin would lie far before recombination and far
before the formation of the first luminous galaxies.

This timing suggests a possible, but highly constrained, connection with the
early structures now being revealed by the James Webb Space Telescope (JWST).
JWST has found luminous and sometimes very massive systems in the first few
hundred million years, including spectroscopically confirmed galaxies at
\(z\simeq14\) and massive galaxies at \(z\simeq5-9\)
\cite{AdamoEtAl2025JWST,CarnianiEtAl2024JWST,XiaoEtAl2024JWST}.  In an astron
interpretation, JWST would not be observing the formation of the charged object
itself.  Rather, it could be observing baryonic structures that form later
around rare pre-existing dark seeds.

The physical picture is therefore that of a non-luminous massive charged source
embedded first in the primordial plasma and later in the intergalactic medium.
The surrounding visible component could be gas, ionization structure, plasma
screening material, or early collapsed baryons.  The compact charged source
would be inferred indirectly through its gravitational field, its plasma
environment, and its lensing properties.

The main obstacle is dynamical.  The net charge corresponds to a tiny fractional
charge imbalance relative to the number of baryons associated with a
\(10^{12}M_\odot\) mass, but the early Universe is a highly conducting plasma.
Any primordial astron scenario must therefore explain both the origin of the
charge imbalance and its survival against neutralization and screening.  If the
charge survives, its cosmological role is still not automatic: the homogeneous
Coulomb energy redshifts as \(a^{-4}\), so any late-time acceleration must be a
transient or domain-dependent effect arising from the inhomogeneous
Einstein--Maxwell problem.

\section{Astron parameters}

We use the following benchmark assumptions.

\begin{enumerate}
\item Astrons are compact enough that their exterior can be treated, at least
outside the material or singular core, by an Einstein--Maxwell electrovac
geometry \cite{Bekenstein1971ChargedFluid,RayEtAl2003ChargedStars,Poisson2004Toolkit}.
\item The fiducial mass scale is ultra-galactic, \(M_A\sim10^{12}M_\odot\).
\item The large-charge branch is the phenomenological charge branch motivated in
Refs.~\cite{Frampton2022EAU,Frampton2023DMDE}, not the ordinary local
accretion-saturation branch.
\item The objects are sparse on cosmological scales, with mean separation of
order a few megaparsecs if their rest mass is not to overclose the Universe.
\item The cosmological question is posed in two stages: first the homogeneous
Friedmann--Lema\^{\i}tre--Robertson--Walker (FLRW) reduction, then the possible
inhomogeneous Einstein--Maxwell backreaction.
\end{enumerate}

This ordering separates three different questions that are sometimes mixed
together:
the microscopic origin of the charge, the compact-object geometry implied by
that charge, and the large-scale cosmological effect of a discrete charged
population.  The separation is essential because the same parameter \(\Xi\)
controls the pairwise force, the Reissner--Nordstr\"om (RN) extremality
threshold and the optical classification of the exterior spacetime.

\section{Relation to previous work}

The astron proposal sits at the intersection of several literatures that are
usually treated separately.  The first is the literature on compact objects
with charge.  In most astrophysical applications the charge of a black hole or
compact star is assumed to be negligible, but there are well-defined settings in
which charge separation, selective accretion and electromagnetic environments
can make the question nontrivial
\cite{Bambi2019,ZajacekTursunov2019,ZajacekEtAl2018,ArayaEtAl2022}.  Charged
fluid spheres and charged compact stars also show that the Einstein--Maxwell
problem is not merely a formal electrovac exercise
\cite{Bekenstein1971ChargedFluid,RayEtAl2003ChargedStars}.

The second is the literature on plasma screening and the intergalactic medium.
The cosmological role of a macroscopic charge cannot be assessed by assigning a
charge to the source and then ignoring the ambient plasma.  The intergalactic
medium (IGM) and circumgalactic medium (CGM) provide the medium through which
electric fields must propagate
\cite{Meiksin2009,McQuinn2016,TumlinsonPeeplesWerk2017}.  Moreover, nonthermal
tails and nonequilibrium plasma distributions can modify the meaning of a
screening length \cite{PierrardLazar2010,LivadiotisMcComas2013,FahrHeyl2016}.
This is why screening is treated here as a central physical constraint.

The third is the literature on optical diagnostics of compact objects.  The RN
and Kerr--Newman metrics belong to the standard family of exact charged
solutions \cite{Kerr1963,NewmanEtAl1965,Carter1968KerrFamily}, and their
lensing properties are controlled by photon orbits and strong-field deflection
coefficients \cite{VirbhadraEllis2000,Bozza2002,EiroaRomeroTorres2002}.  The
astron benchmark lies in a regime where this optical classification changes.
Thus the lensing question is not ancillary to the cosmology; it tests the same
charge-to-mass ratio.

The fourth is the literature on averaging and backreaction.  In homogeneous
FLRW cosmology, the stress-energy content is inserted after averaging.  In a
discrete Einstein--Maxwell universe, the order of operations becomes part of
the problem \cite{Buchert2000,ClarksonEtAl2011,GreenWald2011}.  The astron
scenario therefore connects a charged compact-object problem to a cosmological
coarse-graining problem.  This is the context for the constraints derived below.

\section{Charge generation and saturation}

The first issue is whether an ultra-massive compact object can acquire a
macroscopic charge during its formation.  A minimal capture model shows that the
charge degree of freedom can respond on a timescale much shorter than the
gravitational collapse time.  Charge separation can therefore arise early in the
evolution of the object \cite{ArayaEtAl2022,InayoshiVisbalKashiyama2015,ChonHosokawaYoshida2018,LuoEtAl2018}.

However, the charge does not grow without limit in an ordinary accretion
picture.  For the natural positive branch produced by proton-dominated capture,
the electrostatic potential repels further protons and attracts electrons; the
growth saturates when this feedback suppresses the proton channel or brings the
proton and electron capture rates into balance.  A negative branch would instead
repel electrons.  The resulting charge scale is
parametrically
\begin{equation}
Q_{\rm sat}\sim \frac{GMm_*}{k_e e},
\end{equation}
where \(m_*=m_p\) for the positive proton-limited branch and \(m_*=m_e\) for a
negative electron-limited branch.  This ordinary saturation branch scales as
\(Q_{\rm sat}\propto M\).  It remains far
below the large-charge benchmark when expressed in terms of the geometric
parameter \(\Xi\).
Indeed, inserting the saturation estimate gives
\begin{equation}
\Xi_{\rm sat}
=\frac{k_eQ_{\rm sat}^2}{GM^2}
\sim \frac{Gm_*^2}{k_e e^2},
\end{equation}
which is independent of \(M\) and extremely small for ordinary charged
particles.  This is why local saturation and the large-charge benchmark belong
to different regimes.

The benchmark \(Q_A\sim 4\times 10^{32}\,{\rm C}\) therefore relies on a
separate phenomenological mass-charge extrapolation, rather than on local
accretion saturation.  This is not a small technical distinction.  The ordinary
saturation branch and the large-charge branch correspond to different geometric
sectors.

This distinction also clarifies the relation to charged primordial-black-hole
scenarios.  A primordial object can acquire charge if the capture rates of
positive and negative species differ.  But the sign and magnitude of the charge
depend on the composition of the ambient medium, the velocity distribution, the
finite-size cutoff, and the point at which electrostatic repulsion suppresses
further accretion of the like-charged species.  In such a local description the
charge is self-limiting.  It is therefore misleading to treat the large-charge
benchmark as the inevitable end point of ordinary charge separation.  It is a
separate hypothesis whose implications must be tested directly.

This has an important consequence.  The large-charge branch is interesting
precisely because it is large enough to affect geometry and pairwise dynamics.
But the same largeness makes it harder to justify astrophysically.  The charge
sector is therefore not a solved input to the cosmology.  It is one of the
central assumptions whose microscopic origin must be supplied by the complete
theory.

\begin{figure*}[t]
\centering
\includegraphics[width=0.56\textwidth]{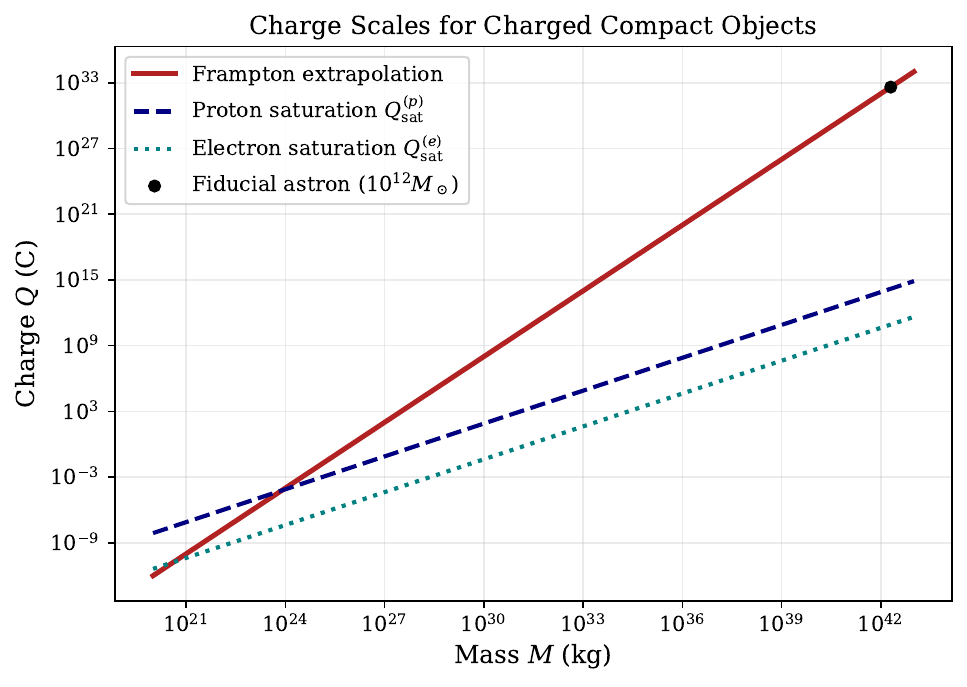}
\caption{Representative charge scales as functions of mass.  The figure
summarizes the separation between ordinary accretion-saturation branches and the
large-charge phenomenological branch.  It is included here only as a compact
statement of the result; the derivation of the curves is deferred to
Ref.~\cite{CorianoEtAlPrep}.}
\label{fig:charge_scales_letter}
\end{figure*}

Figure~\ref{fig:charge_scales_letter} gives a visual summary of this conclusion.
The ordinary saturation branches and the large-charge branch are not nearby
curves in parameter space.  They encode different assumptions about the origin
of the charge.  The remainder of the Letter asks what follows if the
large-charge branch is adopted as a benchmark.

\section{Persistence of charge and plasma screening}

If an astron acquires a large charge, the next question is whether that charge
survives in a cosmological plasma.  Pair creation and local discharge processes
do not automatically remove the charge on cosmological scales for the fiducial
parameters.  The more serious issue is screening by the ionized intergalactic
medium.

The textbook Debye length provides a useful benchmark but not a complete
solution \cite{Meiksin2009,McQuinn2016,DaveEtAl2001,TumlinsonPeeplesWerk2017}.
For a single-temperature electron plasma one would write
\begin{equation}
\lambda_D=
\left(\frac{\epsilon_0 k_BT}{n_e e^2}\right)^{1/2}.
\end{equation}
This linear expression assumes small electrostatic potentials,
near-Maxwellian distributions, and a weak perturbation of the plasma.  Near a
highly charged astron these assumptions are not satisfied.  Nevertheless, the
comparison is instructive: if the relevant screening length is much shorter than
the inter-astron spacing \(L_A\), then long-range electromagnetic correlations
are removed before they can affect cosmology.

Thus a viable cosmological role for the charge requires more than the existence
of a macroscopic \(Q_A\).  It requires a nonlinear and possibly kinetic plasma
description demonstrating that the field remains correlated on scales comparable
to the separation of the sources.

There are two separate plasma questions.  The first is local neutralization:
whether enough opposite charge can be transported into the neighborhood of the
astron to reduce its net charge.  The second is collective screening: whether
the surrounding plasma rearranges so that the electric field is exponentially
suppressed beyond some screening length.  These questions are related but not
identical.  A finite reservoir of opposite charge may exist without immediately
forming a linear Debye cloud, and a linear Debye estimate may fail near the
source even if the far-field tail becomes screened at larger radii.

For cosmology, the second question is decisive.  The astron scenario requires
long-range inter-source correlations.  If the relevant screening length satisfies
\(\lambda_D\ll L_A\), where \(L_A\) is the mean inter-astron separation, then a
charged population reduces to a set of locally screened objects and cannot
generate a collective electromagnetic effect on cosmological scales.  If instead
\(\lambda_D\gtrsim L_A\), the electromagnetic field can correlate many cells,
and the averaging problem becomes intrinsically nonlocal.

The correct treatment is therefore kinetic and dynamical.  It must include the
large source potential, the distribution functions of the plasma species, the
finite time available for rearrangement, and the expansion of the background.
This is why screening appears in our conclusions as a major constraint rather
than a minor environmental correction
\cite{PierrardLazar2010,LivadiotisMcComas2013,FahrHeyl2016,AlonsoMonsalveKaiser2023}.

The screening question also sets the interpretation of any future numerical
work.  A calculation that treats the charge as a fixed external force without
allowing the plasma to rearrange would not answer the cosmological question.  A
calculation that screens the charge locally would test the opposite limit, in
which the electromagnetic sector becomes effectively short-ranged.  The regime
needed for astron cosmology lies between these extremes: the field must be
dynamically self-consistent and still correlated over inter-source distances.

\section{Geometry of highly charged astrons}

Outside a static, spherically symmetric charged object, the electrovac geometry
is of Reissner--Nordstr\"om form,
\begin{equation}
f(r)=1-\frac{2GM}{c^2r}+\frac{Gk_eQ^2}{c^4r^2}.
\end{equation}
Writing
\begin{equation}
m=\frac{GM}{c^2},
\qquad
q^2=\frac{Gk_eQ^2}{c^4}=\Xi m^2,
\end{equation}
the formal RN horizon radii are
\begin{equation}
r_\pm=m\left(1\pm\sqrt{1-\Xi}\right).
\end{equation}
The horizon structure is therefore controlled by \(\Xi\).  Horizons exist only
for \(\Xi\leq 1\).  A large-charge astron with \(\Xi_A\simeq 5.4\) is
super-extremal in the Reissner--Nordstr\"om exterior.

This does not by itself decide whether the object is a singular naked
Reissner--Nordstr\"om spacetime or a horizonless compact object with a regular
interior.  That distinction is an interior-completion question.  What is fixed
by the exterior analysis is the absence of a Reissner--Nordstr\"om horizon for
\(\Xi>1\).

Rotation does not remove the constraint; it enlarges the parameter space.  In a
Kerr--Newman description, charge and angular momentum jointly control the
horizon condition \cite{Kerr1963,NewmanEtAl1965,Carter1968KerrFamily}.  The
static Reissner--Nordstr\"om analysis should therefore be regarded as the
limiting case of a more general charged rotating problem.

The geometrical conclusion concerns the exterior classification.  A realistic
compact object may require a regular interior, a singular interior, a shell-like
completion, or a rotating generalization.  Once the exterior is
Einstein--Maxwell and the charge-to-mass ratio is fixed, however, the exterior
horizon structure is no longer a free assumption.  The large-charge branch is
horizonless in the RN exterior \cite{Israel1966Junction,MazurMottola2004}.

This fact matters phenomenologically.  If the exterior has no horizon, the
object is not simply a massive black hole with an extra force added by hand.  It
belongs to a different sector of the charged compact-object parameter space.
The interpretation of its interior, its stability, and its observational
appearance must be addressed consistently with this exterior classification.

\begin{figure*}[t]
\centering
\includegraphics[width=0.56\textwidth]{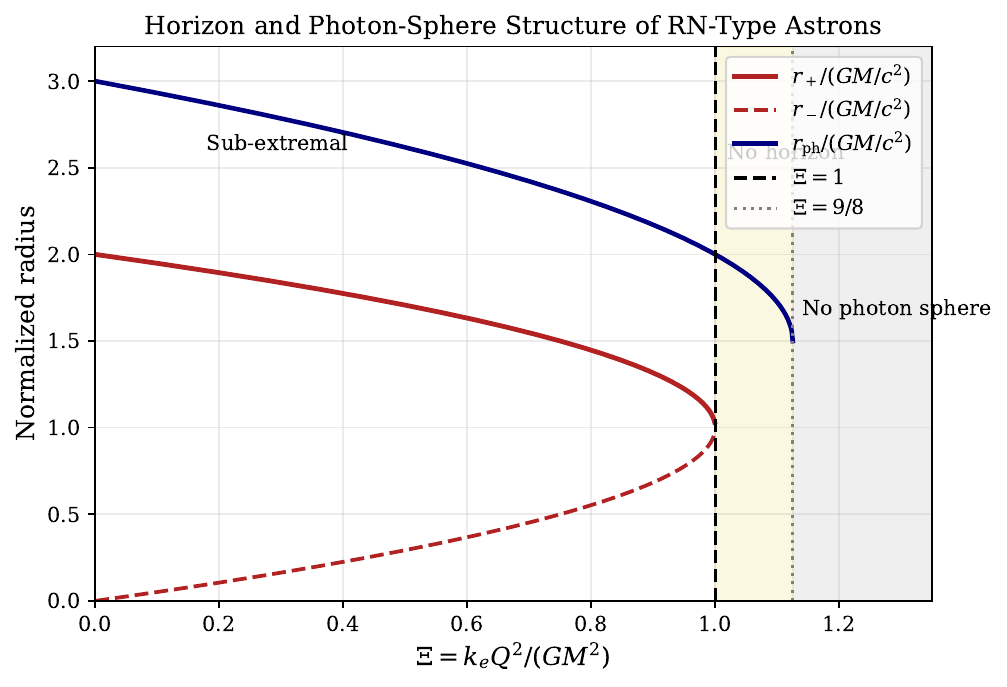}
\caption{RN characteristic radii as functions of \(\Xi\).  The outer horizon
ceases to exist at \(\Xi=1\), while the photon sphere persists only to
\(\Xi=9/8\).  The distinction between these two thresholds is essential for the
interpretation of horizonless but still strongly lensing configurations.}
\label{fig:rn_radii_letter}
\end{figure*}

The existence of two different thresholds is central.  Extremality,
\(\Xi=1\), determines whether a Reissner--Nordstr\"om horizon exists.  The
photon-sphere threshold, \(\Xi=9/8\), determines whether the exterior can still
support black-hole-like strong lensing.  These are close numerically, but they
are not the same physical condition.  The interval between them is therefore a
small but conceptually important regime: horizonless objects in this interval
are not black holes, yet they can retain a photon sphere.

\section{Lensing}

The same geometric parameter controls the optical regime.  Circular null orbits
in a Reissner--Nordstr\"om exterior exist only if
\begin{equation}
\Xi\leq \frac{9}{8}.
\end{equation}
Equivalently, the RN photon radii are
\begin{equation}
r_\gamma^\pm
=\frac{m}{2}\left(3\pm\sqrt{9-8\Xi}\right),
\end{equation}
so real circular null orbits require \(9-8\Xi\geq0\).
The interval \(1<\Xi\leq9/8\) is special: the object is already horizonless, but
it still has a photon sphere and can retain some black-hole-like strong-lensing
features.

For \(\Xi>9/8\), the photon sphere disappears.  This does not mean that the
object becomes optically invisible.  Its mass still produces weak-field
gravitational deflection.  What is lost is the standard black-hole-like
strong-lensing structure associated with photons orbiting the compact object.

The weak-field and strong-field lensing limits have a large literature
\cite{VirbhadraEllis2000,Bozza2002,EiroaRomeroTorres2002,CorianoEtAl2015NeutrinoPhotonLensing,CorianoEtAl2015ElectroweakLensing}.
The weak-field expansion also shows the direction of the charge correction.  For
impact parameter \(b\),
\begin{equation}
\hat\alpha(b)
=\frac{4m}{b}
+\frac{3\pi}{4}(5-\Xi)\frac{m^2}{b^2}
 +O\!\left(\frac{m^3}{b^3}\right).
\end{equation}
At next-to-leading order, the Reissner--Nordstr\"om charge term reduces the
focusing relative to Schwarzschild.  In this precise sense, increasing charge
acts against lensing rather than enhancing it.

This point is useful because it runs counter to a common intuition.  Since the
object is massive, one may expect large lensing.  Since it is also highly
charged, one may expect an even more dramatic optical effect.  The RN geometry
shows a more subtle result.  The mass produces the leading weak-field bending.
The charge affects the subleading structure and, for sufficiently large
\(\Xi\), removes the photon sphere.  The optical signature is therefore not a
stronger black-hole shadow; it is a transition away from the black-hole-like
strong-lensing class.

This provides a diagnostic of the large-charge hypothesis.  A mildly
super-extremal object with \(1<\Xi\le9/8\) is horizonless but still retains a
photon sphere.  A deeply super-extremal object with \(\Xi>9/8\) loses it.  The
benchmark \(\Xi_A\simeq5.4\) lies far into the latter regime.  If objects of
this type existed, their lensing would not be expected to reproduce the
standard sequence of relativistic images associated with black holes of the
same mass.

The lensing result is useful because it is observationally phrased.  The mass of
an astron would still lens light, while the strong-field optical structure is
changed.  If a future model predicts black-hole-like shadows for the
large-charge branch, it must explain how the photon-sphere conclusion is
avoided.

\section{Homogeneous cosmology}

The homogeneous FLRW reduction gives the cleanest no-go statement
\cite{Weinberg2008,DodelsonSchmidt2020,PDG2024Cosmo}.  The astron rest-mass
density scales as ordinary matter,
\begin{equation}
\rho_{M,A}\propto a^{-3}.
\end{equation}
For fixed comoving separation, \(L(a)=aL_0\), the characteristic Coulomb energy
per pair scales as
\begin{equation}
U_C(a)\sim \frac{k_eQ_A^2}{L(a)}\propto a^{-1}.
\end{equation}
Since the number density scales as \(n_A(a)\propto a^{-3}\), the homogeneous
Coulomb interaction energy density scales instead as
\begin{equation}
\rho_C(a)\sim n_A(a)U_C(a)\propto a^{-4}.
\end{equation}
Equivalently, it has the radiation-like equation of state
\begin{equation}
p_C=\frac{1}{3}\rho_C.
\end{equation}
This is also immediate from the continuity equation
\begin{equation}
\dot\rho_C+3H(\rho_C+p_C)=0.
\end{equation}
It therefore contributes positively to \(\rho+3p\) and cannot act as a
cosmological constant.

This result is important because it separates pairwise repulsion from cosmic
acceleration.  A repulsive force between two objects does not automatically
become negative pressure in a homogeneous cosmology.  In the homogeneous
perfect-fluid reduction, the charged interaction behaves like radiation, not
like dark energy.

This is perhaps the most important conceptual point of the investigation.  The
motivation for astrons is often phrased in terms of mutual electric repulsion.
But FLRW acceleration is not controlled by the sign of a pair force.  It is
controlled by the averaged density and pressure that enter the Friedmann
acceleration equation,
\begin{equation}
\frac{\ddot a}{a}
=-\frac{4\pi G}{3}(\rho+3p)+\frac{\Lambda}{3}.
\end{equation}
In a homogeneous reduction, the Coulomb energy of
objects at fixed comoving separation decreases as the universe expands, and its
energy density redshifts like \(a^{-4}\).  This is the same scaling as
radiation, not vacuum energy.

Consequently, the homogeneous astron fluid cannot be used as a substitute for
\(\Lambda\).  If one insists on a homogeneous perfect-fluid description, the
astron charge contributes a positive energy density with positive effective
pressure.  It decelerates rather than accelerates.  This conclusion is robust
within the assumptions of the homogeneous calculation.

The same scaling also fixes the asymptotic interpretation.  A component that
redshifts as \(a^{-4}\) cannot dominate the future expansion in the way a
constant vacuum term does.  Thus even in an effective phenomenological reading
where the charged sector affects the expansion over a finite interval, such an
acceleration era would be transitory rather than asymptotically de Sitter.

There is also a simple abundance constraint.  If one places one
\(10^{12}M_\odot\) object in each cubic megaparsec, the rest-mass abundance is
\begin{equation}
\Omega_A\simeq 7.9,
\end{equation}
which overcloses the Universe.  A matter-like abundance
\(\Omega_A\simeq0.3\) requires a spacing of order
\begin{equation}
L_A\simeq 3\,{\rm Mpc}
\left(\frac{M_A}{10^{12}M_\odot}\right)^{1/3}
\left(\frac{0.3}{\Omega_A}\right)^{1/3}.
\end{equation}

This abundance bound is independent of the electromagnetic sector.  It is simply
the statement that the rest mass of the objects must not exceed the observed
cosmic density budget.  It is also the reason that a literal one-megaparsec
spacing is too dense for \(10^{12}M_\odot\) astrons.  If the population is to
make up only a fraction of the matter density, the spacing must be larger still.

The cosmological role of astrons is therefore doubly constrained.  If the
population is too dense, it overcloses the Universe through rest mass alone.  If
it is sufficiently sparse, the electromagnetic interaction must remain coherent
across larger distances in order to have any collective effect.  The abundance
and screening constraints therefore pull on the same part of the model.

This point also shows why the cosmological problem cannot be reduced to a single
parameter.  Increasing the number density makes the population more
cosmologically important, but it also increases the rest-mass abundance.
Increasing the separation avoids overclosure, but it requires electromagnetic
correlations to survive over larger distances.  Increasing the charge strengthens
pairwise repulsion, but it pushes the exterior geometry deeper into the
super-extremal and no-photon-sphere regime.  The model is constrained by the
simultaneous intersection of these requirements.

\section{Backreaction as the remaining loophole}

The failure of the homogeneous approximation does not exclude a genuinely
discrete Einstein--Maxwell cosmology.  It identifies the place where the next
calculation must be done.  The homogeneous procedure averages the system first
and evolves the resulting FLRW model.  A more faithful treatment evolves the
inhomogeneous system and only then averages the result.

Because Einstein's equations are nonlinear, these operations need not commute.
This is the origin of cosmological backreaction
\cite{Buchert2000,ClarksonEtAl2011,GreenWald2011}.  In a spatial domain
\({\cal D}\), define
\begin{equation}
V_{\cal D}(t)=\int_{\cal D}\sqrt{h}\,d^3x,
\qquad
\langle S\rangle_{\cal D}
=
\frac{1}{V_{\cal D}}\int_{\cal D}S\sqrt{h}\,d^3x ,
\end{equation}
where \(h\) is the determinant of the induced spatial metric.  The associated
domain scale factor is
\begin{equation}
a_{\cal D}(t)=
\left[\frac{V_{\cal D}(t)}{V_{\cal D}(t_0)}\right]^{1/3}.
\end{equation}
Following Buchert's irrotational-dust average \cite{Buchert2000}, we define the
Buchert kinematical backreaction term \(Q_{\cal D}\) by
\begin{equation}
\label{eq:letter_QD_definition}
Q_{\cal D}
\equiv
\frac{2}{3}
\left(
\langle\theta^2\rangle_{\cal D}
-\langle\theta\rangle_{\cal D}^2
\right)
-2\langle\sigma^2\rangle_{\cal D},
\end{equation}
with \(\theta=\nabla_\mu u^\mu\) the local expansion and
\(\sigma^2=\sigma_{\mu\nu}\sigma^{\mu\nu}/2\) the shear scalar.  For charged
sources the congruence is not generally geodesic.  We denote the corresponding
Lorentz-force acceleration contribution \(A_{\cal D}\) by
\begin{equation}
\label{eq:letter_AD_definition}
A_{\cal D}
\equiv
\left\langle\nabla_\mu a^\mu\right\rangle_{\cal D},
\qquad
a^\mu=u^\nu\nabla_\nu u^\mu
\simeq
\frac{\rho_q}{\rho_M}F^\mu{}_\nu u^\nu ,
\end{equation}
for a charged dust element, with \(\rho_q\) and \(\rho_M\) the charge and mass
densities.  Thus \(A_{\cal D}\) is not part of the standard Buchert dust system;
it is the extra Einstein--Maxwell, boundary-sensitive term.

The corresponding Einstein--Maxwell acceleration diagnostic is
\begin{equation}
3\frac{\ddot a_{\cal D}}{a_{\cal D}}
=
-4\pi G\langle\rho_M\rangle_{\cal D}
-8\pi G\langle\rho_{\rm EM}\rangle_{\cal D}
+Q_{\cal D}
+A_{\cal D}
+\Lambda .
\end{equation}
Here \(\rho_M\) and \(\rho_{\rm EM}\) denote the matter and electromagnetic
energy densities. 
Thus, without a fundamental cosmological constant, acceleration requires
\begin{equation}
Q_{\cal D}+A_{\cal D}
>
4\pi G\langle\rho_M\rangle_{\cal D}
+8\pi G\langle\rho_{\rm EM}\rangle_{\cal D}.
\end{equation}
For an astron abundance \(\Omega_A\), this implies the target scale
\begin{equation}
\frac{Q_{\cal D}+A_{\cal D}}{H_0^2}
\gtrsim
\frac{3}{2}\Omega_A.
\end{equation}
If \(\Omega_A\simeq0.3\), the required positive contribution is of order
\(0.5H_0^2\).  This is not a small correction.

Backreaction is therefore not a conclusion; it is a precise open problem.  A
successful astron cosmology must derive a positive contribution of this size from
the discrete Einstein--Maxwell dynamics, while avoiding cancellation by shear,
Maxwell focusing, finite-domain effects and plasma screening.

This statement also explains why the word ``backreaction'' is used.  In the
FLRW treatment one averages first and evolves the resulting homogeneous model.
In the backreaction treatment one evolves the inhomogeneous system and only then
averages.  The difference between these two operations is the feedback of
small-scale structure on the effective large-scale expansion.  For astrons, the
small-scale structure is not merely a density contrast.  It includes compact
charged sources, anisotropic electromagnetic stress and directed Lorentz forces.
Exact charged black-hole cosmologies provide useful comparisons, although they
do not by themselves solve the discrete astron averaging problem
\cite{BibiCliftonDurk2017}.

With Eqs.~\eqref{eq:letter_QD_definition} and
\eqref{eq:letter_AD_definition}, \(Q_{\cal D}\) is the usual Buchert
kinematical backreaction, while \(A_{\cal D}\) is the charged-source addition.
Neither term is guaranteed to be positive, and neither term is guaranteed to be
large.  This is why the backreaction route is a loophole in the homogeneous
no-go result, but not a demonstrated solution.

\begin{figure*}[t]
\centering
\includegraphics[width=0.56\textwidth]{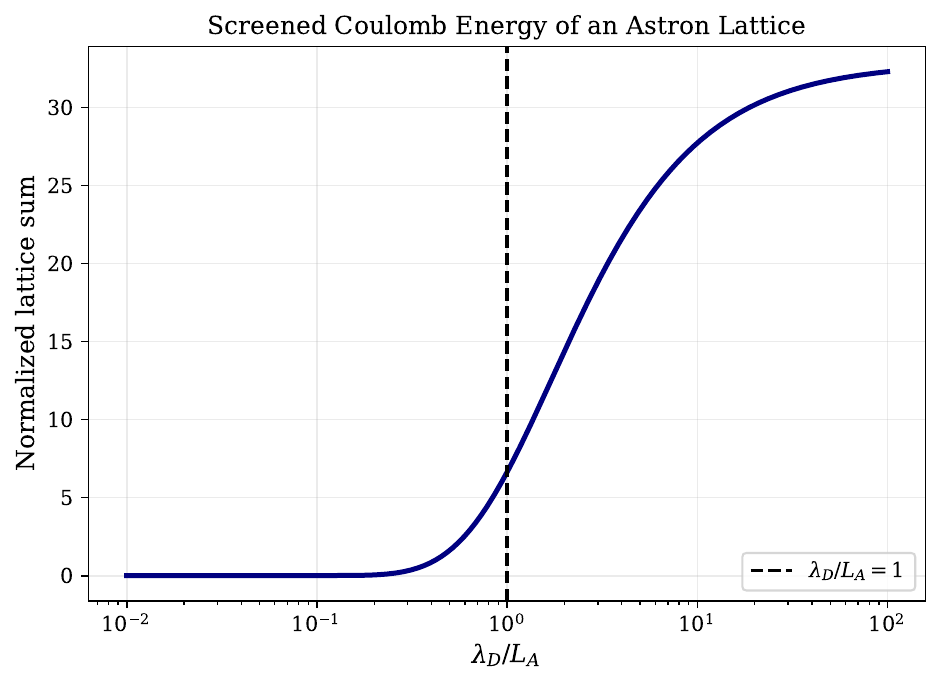}
\caption{Screened lattice diagnostic for a Yukawa-regulated inter-astron
interaction.  If the screening length is shorter than the inter-source spacing,
the collective electromagnetic contribution is suppressed.  Long-range
correlations survive only when the screening scale is comparable to or larger
than the spacing.}
\label{fig:screened_lattice_letter}
\end{figure*}

Figure~\ref{fig:screened_lattice_letter} is included as a diagnostic for the
remaining acceleration problem.  It shows that the screening and backreaction
questions are inseparable.  If screening is local, the backreaction problem
collapses back toward the homogeneous or short-range result.  If screening is
weak on the inter-astron scale, the electromagnetic field remains nonlocal and a
true Einstein--Maxwell averaging calculation is required.

\section{Synthesis of the constraints}

The preceding results should be read together, because the same parameters are
tested in several different ways.  The charge required for strong pairwise
repulsion is also the charge that determines the Reissner--Nordstr\"om
extremality parameter \(\Xi\).  The separation required to avoid overclosure is
also the scale over which the electric field must avoid efficient plasma
screening.  The electromagnetic field that can influence the averaged expansion
also carries positive energy density and anisotropic stress.  The model is
therefore constrained by a simultaneous intersection of charge generation,
screening, geometry, lensing, abundance and averaging.

The large-charge branch is not the ordinary local saturation branch.  Minimal
capture and saturation give \(Q_{\rm sat}\propto M\), but
\(\Xi_{\rm sat}\ll 1\) for ordinary charged particles.  The benchmark value
\(\Xi_A\simeq5.4\) must therefore be supplied by an additional formation
mechanism or phenomenological mass-charge relation.  Once this benchmark is
adopted, however, the exterior classification is fixed: the RN horizons are
absent for \(\Xi>1\), and the photon sphere is absent for \(\Xi>9/8\).  The
fiducial branch lies in the horizonless, no-photon-sphere sector.

The cosmological constraints are equally direct.  The homogeneous rest-mass
density of the population scales as \(a^{-3}\), while the homogeneous Coulomb
energy scales as \(a^{-4}\) and has \(p_C=\rho_C/3\).  Thus the homogeneous
charged component is radiation-like, not vacuum-like.  Moreover, a literal
one-megaparsec spacing for \(10^{12}M_\odot\) astrons overcloses the Universe;
a matter-like abundance requires separations of order several megaparsecs.  A
cosmological role for the charge therefore requires both a sparse population and
electromagnetic correlations that survive over the corresponding distances.

Backreaction is the remaining route beyond this homogeneous result.  It is not a
replacement name for the Coulomb energy.  It denotes the difference between
averaging a lumpy Einstein--Maxwell system before evolution and averaging it
after evolution.  In the notation above, the relevant domain source is the
combination \(Q_{\cal D}+A_{\cal D}\), where \(Q_{\cal D}\) is the Buchert
kinematical term and \(A_{\cal D}\) is the charged-source acceleration term.
For an astron abundance \(\Omega_A\), acceleration requires a contribution of
order
\begin{equation}
\frac{Q_{\cal D}+A_{\cal D}}{H_0^2}
\gtrsim
\frac32\Omega_A .
\end{equation}
For \(\Omega_A\simeq0.3\), this is an \(O(H_0^2)\) effect, not a perturbative
correction that can be assumed to appear automatically.

\section{Observable and cosmological tests}

The first observational test is optical.  A large-charge astron still produces
weak-field lensing through its mass, but the benchmark RN exterior lacks the
photon sphere that generates the standard black-hole-like sequence of
strong-field relativistic images.  The lensing signal would therefore separate
the mass scale of the object from the optical class of the exterior spacetime.

The second test is environmental.  The intergalactic medium and circumgalactic
medium are part of the dynamical system through which the electric field must
propagate.  A linear Debye length is only a diagnostic near a highly charged
source, but the implication is sharp: efficient screening on scales shorter than
the mean separation erases the long-range electromagnetic sector.  Weak
screening on inter-source scales instead turns the problem into a genuinely
nonlocal Einstein--Maxwell averaging problem.

The third test is cosmological.  An effective background description could be
written schematically in terms of the normalized expansion rate
\(E(a)\equiv H(a)/H_0\) as
\begin{equation}
E^2(a)=
\Omega_{r0}a^{-4}
+\Omega_{m0}a^{-3}
+\Omega_{A0}X_A(a),
\end{equation}
but \(X_A(a)\) cannot be identified with the homogeneous Coulomb energy.  It
would be a normalized effective astron contribution, \(X_A(1)=1\), summarizing
discreteness, screening, Lorentz-force acceleration,
anisotropic Maxwell stress and domain dependence.  Its sign, magnitude and time
dependence must be derived from the discrete Einstein--Maxwell problem before
comparison with \(H(z)\), distances, growth of structure and lensing data.

These statements define the next tests of the scenario.  One needs a formation
history for the large-charge branch, a nonlinear or kinetic treatment of plasma
screening, a consistent completion of the super-extremal exterior, and a domain
averaging calculation in which the kinematical backreaction, Maxwell stress,
Lorentz-force acceleration and boundary terms are all kept under control.  A
small or negative value of \(Q_{\cal D}+A_{\cal D}\) leaves the homogeneous
no-acceleration result intact.  A positive value of order \(H_0^2\), together
with weak screening and acceptable lensing phenomenology, would identify the
regime in which astrons could act as a genuine Einstein--Maxwell alternative to
a homogeneous dark-energy fluid.

\section{Conclusion}

Astrons provide a sharply constrained framework for connecting compact charged
objects to large-scale cosmology.  Ordinary charge saturation is far below the
large-charge benchmark; the benchmark branch is super-extremal and lacks a
photon sphere; plasma screening can remove the long-range force; and the
homogeneous Coulomb component redshifts like radiation rather than vacuum
energy.

The physical picture is therefore not that of a luminous object, nor that of a
black hole whose lensing is simply strengthened by charge.  An astron would be a
dark massive charged source embedded in a dilute plasma environment.  Its mass
can still produce weak lensing, but the charge reduces the Reissner--Nordstr\"om
strong-lensing structure and can eliminate the photon sphere.  This is the main
shift relative to the initial phenomenological picture based only on mutual
electrostatic repulsion: the same charge that gives a large pairwise force also
changes the compact-object optics, the plasma problem and the cosmological
averaging problem.

If astrons are related to the early structures observed by JWST, the connection
would be indirect: the charged objects would act as primordial dark seeds, while
the telescope would observe the later baryonic systems assembled around them.

The cosmological viability of the scenario therefore rests on a derived
backreaction effect in a sparse, screened, discrete Einstein--Maxwell system.
This is the calculation that must replace the homogeneous perfect-fluid
reduction.  Details of the capture model, saturation estimates, screening
analysis, compact-object classification, lensing calculation and averaging
formalism will be presented in the longer work in preparation
\cite{CorianoEtAlPrep}.

\section*{Acknowledgements}

This work is supported by Iniziativa Specifica INFN \emph{QG-SKY}.

\end{document}